\documentclass{PoS}

\title{On curing the divergences in the quark number susceptibility}

\ShortTitle{Curing Divergences}

\author{\speaker{Rajiv V.\ Gavai}\thanks{Sir J.C. Bose National Fellow.}\\
         Department of Theoretical Physics, Tata Institute of Fundamental
         Research,\\ Homi Bhabha Road, Mumbai 400005, India.  \\
        E-mail: \email{gavai@tifr.res.in}}

\author{Sayantan\ Sharma\thanks{Currently at Physics Department, Brookhaven
National Laboratory, Upton, NY11973}\\
        Fakult\"at f\"ur Physik, Universit\"at Bielefeld,\\
        D-33615 Bielefeld, Germany\\   
        E-mail: \email{sayantan@physik.uni-bielefeld.de}}

\abstract{Adding chemical potential $\mu$ linearly as $\mu N$ to the lattice QCD
action, where $N$ is a conserved quark/baryon number, leads to a quadratic 
divergence as $a^{-2}$.  We argue that it is inherited from the continuum 
theory and can be subtracted off on the lattice following a similar manner in
the continuum.  We test this idea for quenched quark number susceptibilities and
demonstrate a finite continuum limit numerically.  }

\FullConference{The 32nd International Symposium on Lattice Field Theory,\\
		23-28 June, 2014\\
		Columbia University New York, NY}

\begin{document}

\section{Introduction}

The quark number susceptibilities are important measurables on the lattice due
to their relation to the fluctuations of net charge or baryon number that are
being measured in the heavy ion collision experiments at RHIC at Brookhaven
National Laboratory. Apart from providing information about the degrees of
freedom and the interactions of the hot and dense QCD medium, these are also
important in estimating the location of the critical point in the QCD phase
diagram.  The critical point is defined by the singularity of the baryon number
susceptibility.  Using successive derivatives of the pressure with $\mu_B/T$,
its location maybe determined by the radius of convergence of the Taylor
expansion for it.  The radius of convergence depends on the ratios of the higher order
quark number susceptibilities, so their accurate and efficient measurement is a
challenging problem on the lattice. In order to look for an economic method of
computing them, we revisit the problem of introducing chemical potential on the
lattice.

As in the continuum, the canonical and most simple way of introducing chemical
potential, $\mu$ is by adding a term $\mu N$ to the QCD Lagrangian where $N$ is
the charge corresponding to the conserved point-split current on the lattice. It
amounts to introducing weights $f(a \mu) = 1 + a \mu$ \& $g(a \mu) = 1 - a \mu$
to forward and backward time links in the Dirac operator respectively. However
this leads to $\mu$-dependent $a^{-2}$ divergences in energy density and quark
number density. To cure this on the lattice one of the popular
method~\cite{hk,kogut} is to modify the weights to $\exp (\pm a \mu)$ to obtain
finite results.  This proposal is not unique and weights $( 1 \pm a \mu)/\sqrt (
1- a^2 \mu^2)$ for the temporal links also lead to finite results~\cite{bilgav}.
In fact, all that is needed is that any general weights $f,g$ should follow $
f(a\mu) \cdot g(a \mu) =1$ with $f(0) = f'(0) =1$ in order to cure the undesired
divergences on the lattice~\cite{gavai}. It is worth emphasizing here that the
analytical proof in all the cases above was for free quarks. Indeed, further
numerical computations in quenched QCD showed it to work for the interacting
case as well~\cite{ggquen}, while a similar check in the full theory is still
lacking.

Due to the remnant chiral symmetry the staggered quarks possess, a majority of
the numerical computations have so far been performed employing them; the chiral
symmetry issue gets even more complex for the Wilson fermions.   As is
well-known, the Overlap and/or Domain Wall Fermions are much more preferable
from the chiral symmetry perspective. They have {\em both} the correct chiral
and flavour symmetry on lattice as well as an index theorem on the lattice
\cite{hln,Lues}.  These are likely to be crucial for investigations of the QCD
critical point, which model-based considerations show to result if QCD has only
two light quarks and the axial $U_A(1)$ symmetry-restoration takes place at
sufficiently large temperatures.  A moderately heavy strange quark may affect
the location of the QCD critical point but not its existence.

Non-locality of the overlap fermions makes the introduction of the chemical
potential as a Lagrange multiplier of the conserved charge nontrivial. Bloch and
Wettig \cite{wettig} proposed to use the same prescription as above for the
timelike links of the Wilson-Dirac kernel $D_W(a \mu)$ to define the
corresponding overlap Dirac matrix at nonzero density.  The resultant overlap
fermion action indeed has no $a^{-2}$ divergences \cite{gl,bgs} in the free
case.  Unfortunately, however, it has no chiral invariance for nonzero $\mu$
either \cite{bgs}.  Using the definition of the chiral projectors for overlap
fermions, we \cite{gsov} proposed a chirally invariant Overlap action for
nonzero $\mu$ : 
\begin{eqnarray} 
\nonumber
S^F &=& \sum_n  [\bar \psi_{n,L} (aD_{ov} + a\mu
\gamma^4) \psi_{n,L}+\bar \psi_{n,R} (aD_{ov} + a\mu \gamma^4) \psi_{n,R}]  \\
\nonumber &=& \sum_n  \bar \psi_n [ aD_{ov} + a\mu \gamma^4
( 1- aD_{ov}/2) ] \psi_n  ~. 
\end{eqnarray}

It is easy to check that 
\begin{itemize}

\item the action is indeed invariant under the chiral transformations, 
$\delta \psi = i \alpha  \gamma_5(1 - a D_{ov}) \psi$ and $\delta \bar \psi 
=i \alpha \bar \psi \gamma_5$, for any value of $a\mu$ or $a$.

\item it reproduces the continuum action in the limit $a \to 0$ under 
$a\mu \to a\mu/M$ scaling, $M$ being the irrelevant parameter in overlap action.

\item the chiral order parameter exists for all $\mu$ and $T$.  It is defined
by 
$$\langle \bar \psi \psi \rangle = 
\lim_{am \to 0}~~ \lim_{V \to \infty}  \langle {\rm Tr~} {\frac{(1-a D_{ov}/2)} 
{[ a D_{ov} + (am + a\mu \gamma^4) (1 -a D_{ov}/2)]} } \rangle.$$

\end{itemize}

It, however, has the same $\mu$-dependent $a^{-2}$ divergences in the number density and 
the energy density as the linear $\mu$-case for naive/staggered fermions \cite{ns}.   
Furthermore, unlike for the latter fermions, these cannot be removed \cite{ns} by 
exponentiation of the $\mu$-term.  This leads us to a dilemma in case of the Overlap 
fermions: one either one sacrifices exact chiral invariance on the 
lattice or have it with the divergences in the continuum limit of $a \to 0$.

\section{Tackling the Divergences }

Motivated thus by the desire to maintain exact chiral symmetry at a finite
lattice spacing for any value of the chemical potential $\mu$ in order to
have a well defined order parameter protected by the chiral symmetry as a
function of $(\mu,T)$, we opted to explore whether the $\mu$-dependent $a^{-2}$
divergences can be tackled without adopting the popular route of change of 
the fermionic action using terms that vanish in the continuum limit.

As a first step we examined carefully the free dense quark gas in continuum.
We found that contrary to the common belief, these divergences are {\em not} 
due to lattice artifacts.   Indeed, the $\mu$-dependent divergences exist in the
continuum theory as well when appropriate care is taken while manipulating divergent
integrals. The lattice regulator simply makes it easy to spot them.  Using
a Pauli-Villars cut-off $\Lambda$ in the continuum theory, one can also show 
the presence of $\mu \Lambda^2$ term in number density easily \cite{gs14}.

We will sketch the argument here briefly, referring the reader to \cite{gs14}
for more details.  The expression for the free quark number density can be
easily derived to be
\begin{equation}
\nonumber
 n=\frac{2iT}{ V}\sum_{n}\int \frac{d^3p}{(2\pi)^3}\frac{(\omega_n-i\mu )}
 {p^2+(\omega_n-i\mu )^2}~ \equiv \frac{2iT}{ V}\sum_{n}\int \frac{d^3p}{(2\pi)^3} F(\omega_n,\mu ,\vec p),
\nonumber
\end{equation}
where $p^2= p_1^2+ p_2^2+ p_3^2$.  All the gamma matrices are
Hermitian in our convention.  $T=0$ and $\mu=0$ corresponds to the vacuum 
contribution which can be removed by subtracting $n(T=0, \mu=0)$.  While this
is identically zero, the corresponding subtraction for the energy density is
actually $ \propto \Lambda^4$.

\begin{figure}
\begin{center}
\includegraphics[scale=0.4]{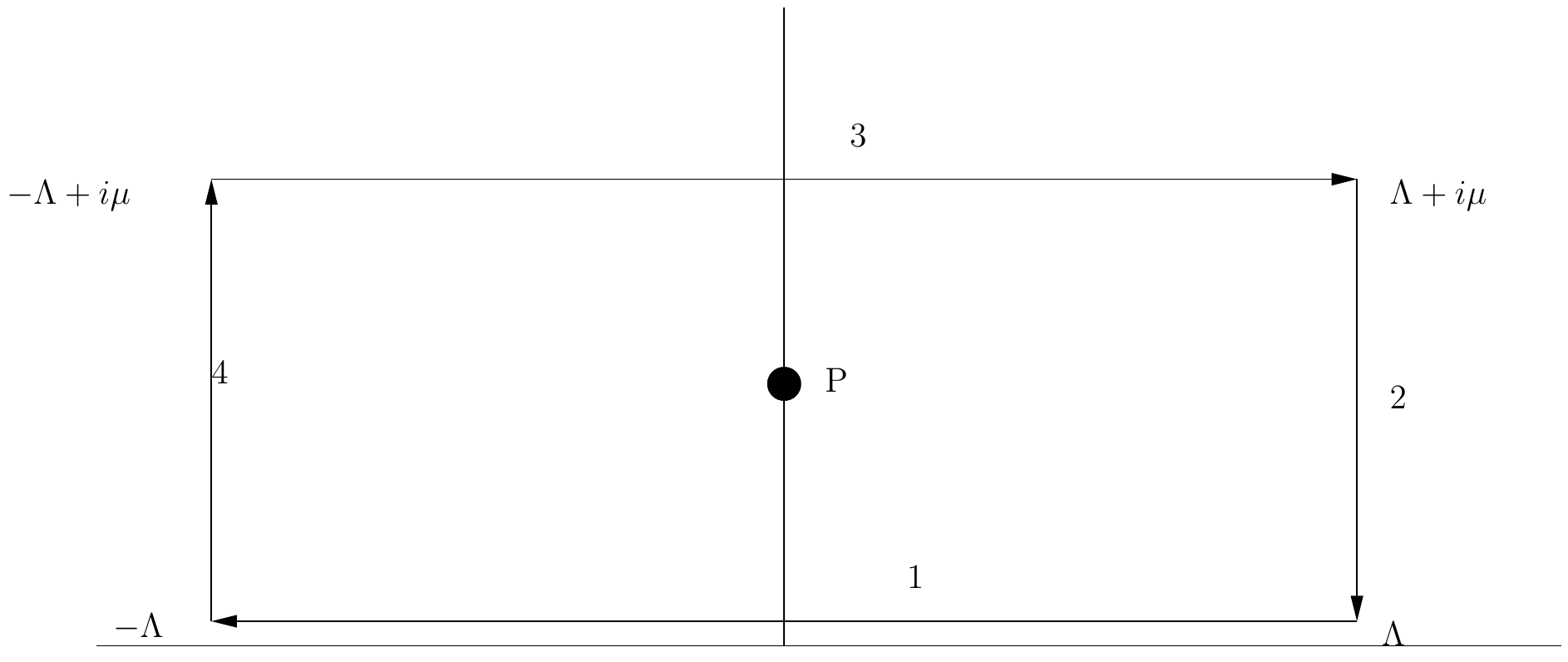}
\caption{The contour diagram used for calculating the number density for free 
fermions in the continuum.}
\label{fig:contour}
\end{center}
\end{figure}

Employing the usual contour method, but with a cut-off $\Lambda$ for all four
momenta, one has in the $T \to 0$ limit but $\mu \ne 0$ the number density 
as,
\begin{equation}
\label{eqn:ncont2}
 n = 2i\int\frac{d^3p}{(2\pi)^3}\left[-i \Theta
\left(\mu- p \right) -\Bigg( \int_2+\int_4+\int_1 \Bigg) \frac{d \omega}{\pi} 
\frac{\omega} {p^2 +\omega^2} \right]~.~
\end{equation}
The corresponding contour diagram is shown in Figure \ref{fig:contour}.

Clearly the contribution of the arms 2 and 4 in the figure arises as extra terms
due to the completion of the contour. The $\mu \Lambda^2$ terms can be shown to
arise from the terms coming coming from these arms \cite{gs14}, being of the
form log $\left[ \frac{p^2+(\Lambda+i \mu)^2}{p^2+(\Lambda-i \mu)^2}\right]$.
If one lets $\Lambda \to \infty$, one may na{\i}vely set the term to zero.  A
careful expansion in $\mu/\Lambda$ shows, however, that while the leading
$\Lambda^3$ terms do cancel from the numerator and denominator, the
$\mu\Lambda^2$ terms add together, and survive.  Dropping the contribution from
the arms 2 and 4, as is usually done, amounts to a  subtracting the divergence
by hand.  One may follow this prescription of subtracting the free theory
divergence by hand on the lattice as well.  If it works, one can not only
explore the QCD phase diagram with exact chiral fermions, but also have several
computational advantages in computing the higher order susceptibilities needed
in critical point search.

Indeed, for any fermion it leads to $M' = \sum_{x,y} N(x,y)$, and $ M'' = M''' =
M''''... = 0$, in contrast to the exp($ \pm a \mu$)-prescription where {\em all}
derivatives are nonzero: $M' , M'''... \ne 0 $ and $ M'' , M'''' , M''''''...\ne
0$. As pointed out in~\cite{gs10}, this results in a  lot fewer terms in each of 
the Taylor coefficients, especially as the order increases.   E.g., in the 4th 
order susceptibility, the dominant contribution to the diagonal part is due to the 
operator $\mathcal{O}_4 = - 6 ~{\rm Tr ~} (M^{-1}M')^4$ which has only a single 
term in the linear case compared to $\mathcal{O}_4 = - 6~ {\rm Tr ~}(M^{-1}M')^4 + 
12 ~{\rm Tr ~} (M^{-1}M')^2 M^{-1}M'' -3 ~{\rm  Tr ~}
(M^{-1}M'')^2 - 3 ~{\rm Tr ~} M^{-1}M'M^{-1}M''' + {\rm Tr ~} M^{-1}M''''$ in
the exponential( or generic $f \dot g =1)$ case.  The eighth order coefficient
has a major contribution from $\mathcal{O}_8$.  It too has one term for the 
linear case. To compute it one has to do eight sequential matrix inversions in contrast to 
18 in the usual case.  Moreover unlike in the linear method the $\mathcal{O}_n$s have 
terms that occur with varying signs, these usually
give rise to large fluctuations, and thus huge error bars, in determining the
higher coefficients.  Less cancellations may lead to better error estimate for
the same CPU time and lesser number of $M^{-1}$ computations needed reduced the
time even proportionately.

\section{Testing the idea }

\begin{figure}
\begin{center}
\includegraphics[scale=0.5]{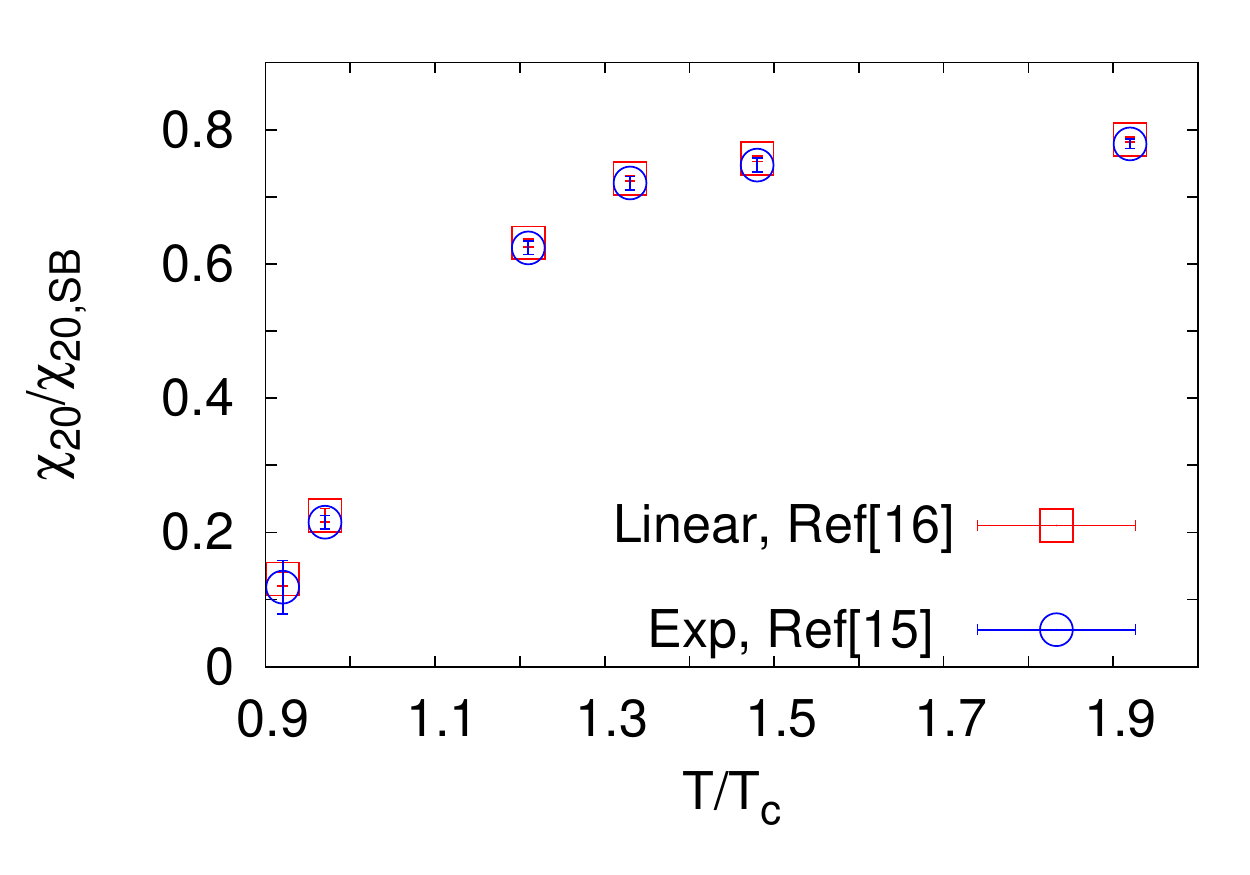}
\caption{The ratio of the second order quark number susceptibility $\chi_{20}$ for two flavour QCD 
divided by the corresponding Stefan-Boltzmann value computed on the lattice of same size, compared 
at different temperatures for the linear and the exponential $\mu$ method(GG).}
\label{fig:chi2bchi2sb}
\end{center}
\end{figure}

We had earlier tested this proposal in the linear $\mu$ case with a simple
subtraction of the free gas result on the $N_t=6$ full QCD ($N_f=2$)
configurations\cite{gg1} by computing\cite{gs12} all the coefficients and
comparing them to the published results with the action exponential in $\mu$,
i.e., without any subtraction. A good agreement within errors was observed.  The
Figure~\ref{fig:chi2bchi2sb} displays the most precisely computed first
coefficient $\chi_{20}$ normalized by its Stefan-Boltzmann value computed on the
lattice of same dimensions. It is natural to expect different finite size
effects for different actions hence the results were normalized with the
corresponding free fermion results for the $N_t=6$ case.  

However, such a comparison falls short of testing the presence of any residual
effects due to inadequate subtraction or even overcompensation since it is made
at just one lattice spacing at each temperature.  In order to test whether the
divergence is truly absent, one needs to take the continuum limit $a \to 0$ or
equivalently $N_t \to \infty$ at a fixed physical scale or temperature.
Carrying out such a program for the full QCD is currently computationally hard.
On the other hand, simulations for quenched QCD are easier done for larger $N_t$
needed for such an exercise.  Considering the known empirical agreement for
dimensionless ratios of physical quantities computed in quenched QCD and full
QCD, such as the hadron spectrum or the pressure ($p/T^4$) as a function of
$T/T_c$, one expects the test in quenched QCD to be a good indication, in case
the divergent effects are changed by interactions. 

\begin{figure}[htb]
\begin{center}
\includegraphics[scale=0.475]{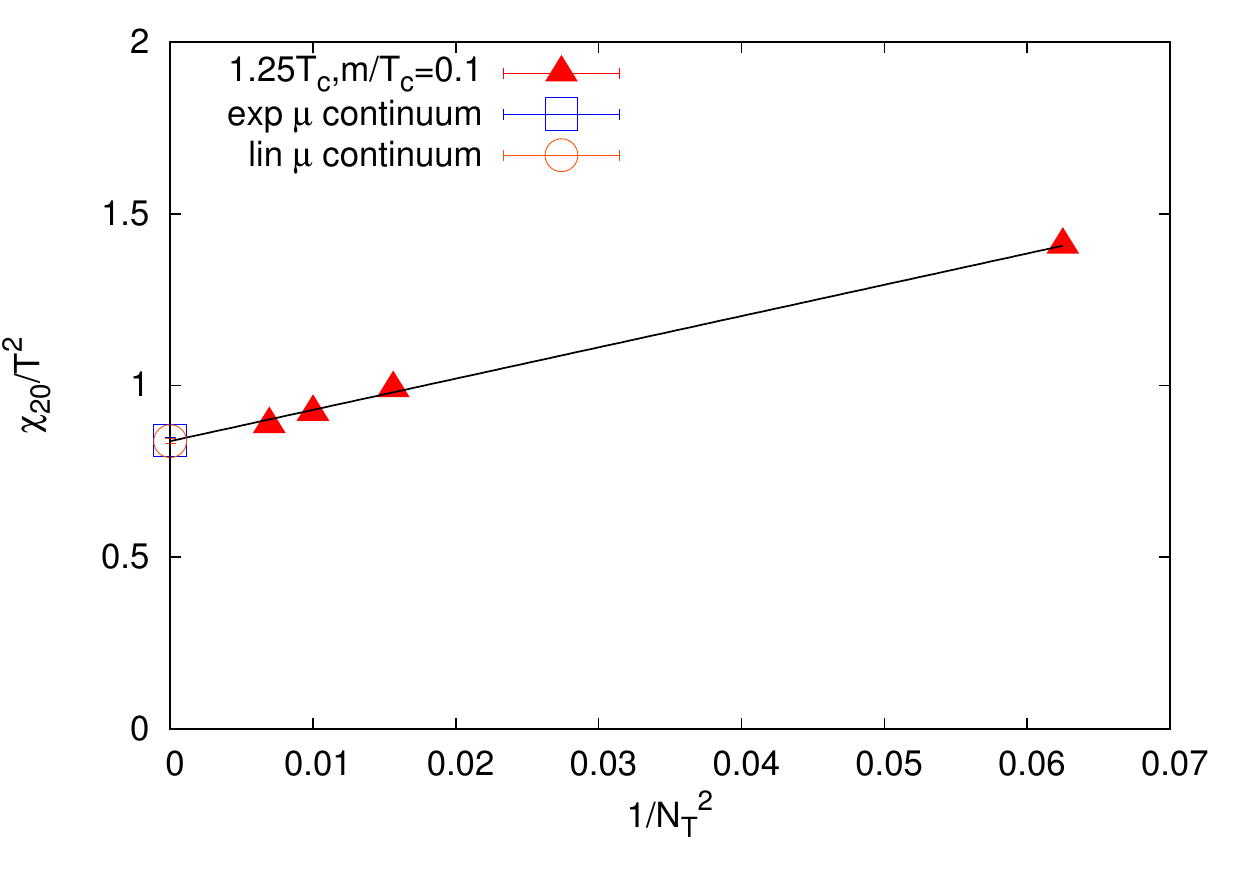}
\includegraphics[scale=0.475]{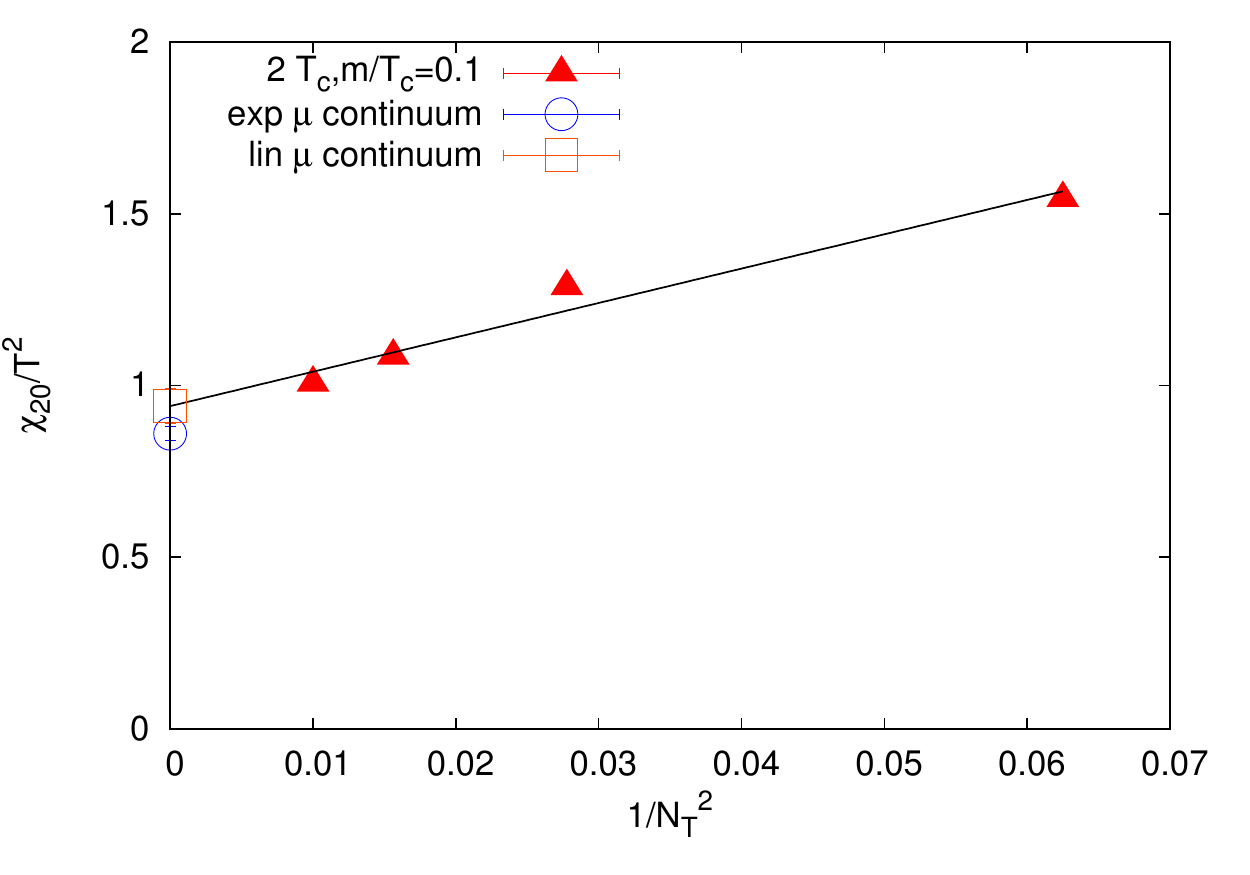}
\caption{The continuum extrapolation for the second order susceptibility in
quenched QCD at $1.25 T_c$ (left panel) and $2 T_c$ (right panel) as a function
of the lattice spacing or equivalently $N_t$ for $m/T_c=0.1$. The blue squares
are the continuum results from the conventional method taken from
Ref.~\cite{swagato}.}
\label{fig:chi2quench1}
\end{center}
\end{figure}

We therefore tested the idea for quenched QCD by employing $N_t = 4, 6, 8, 10$
and $12$ lattices at two different temperatures, $T/T_c = 1.25$ and $2$.  These
were chosen to exploit the known results for the critical couplings as a
function of $N_t$ to avoid any fine-tuning in scale fixing.  The results
displayed below are for valence quark mass of $m/T_c =0.1$.   The quark number
susceptibility was measured on typically 50-100 independent configurations.  The
subtraction of the free gas divergent term was done for the respective $N_t$
in each case with the same values of valence quark mass.  Absence of any
divergent term in either case is evident from the positive slope of the data at
both temperatures shown in Figure~\ref{fig:chi2quench1}.  Furthermore, the
extrapolated continuum result coincides with the earlier result obtained with
the $\exp(\pm a \mu)$ action \cite{swagato}. 

 Since the valence quark mass introduces yet another scale a divergence, if
present, could arise while reducing it.  We lowered the mass by a factor of 10
to $m/T_c =0.01$ and repeated the exercise at the lower temperature, 
$T/T_c =1.25$.
\begin{figure}
\begin{center}
\includegraphics[scale=0.5]{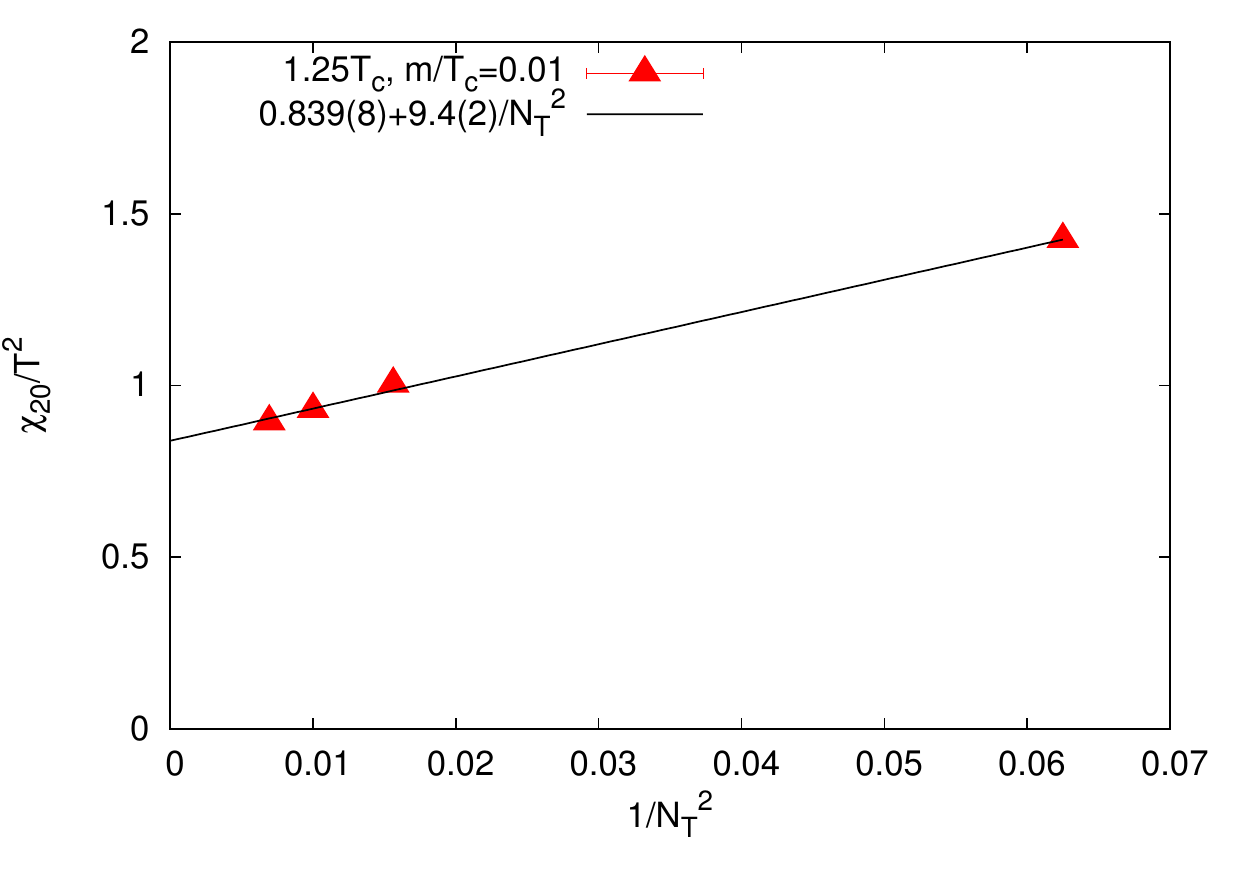}
\caption{The continuum extrapolation for the second order susceptibility data in quenched QCD at 
$1.25 T_c$ for $m/T_c=0.01$.}
\label{fig:chi2quench2}
\end{center}
\end{figure} 
Once again the trend of the data in Figure~\ref{fig:chi2quench2} is same, and no
divergent term is evidently present as inferred from the slope of the data.  The
comparison with the continuum extrapolated result with the exponential method
was not feasible as the corresponding result for that action did not exist.

Finally, as a check of the scheme we also studied the next higher order
susceptibility in a similar way.  The fourth order susceptibility would have an
additional unphysical $\mathcal{O}(a^0)$ term which we remove by subtracting the
zero temperature part of the corresponding free fermions.  It too shows similar
finite result in continuum limit without any hints of any upward trend
indicative of divergences.  Moreover, a reasonable continuum limit is also
visible by extrapolation.

\begin{figure}[htb]
\begin{center}
\includegraphics[scale=0.425]{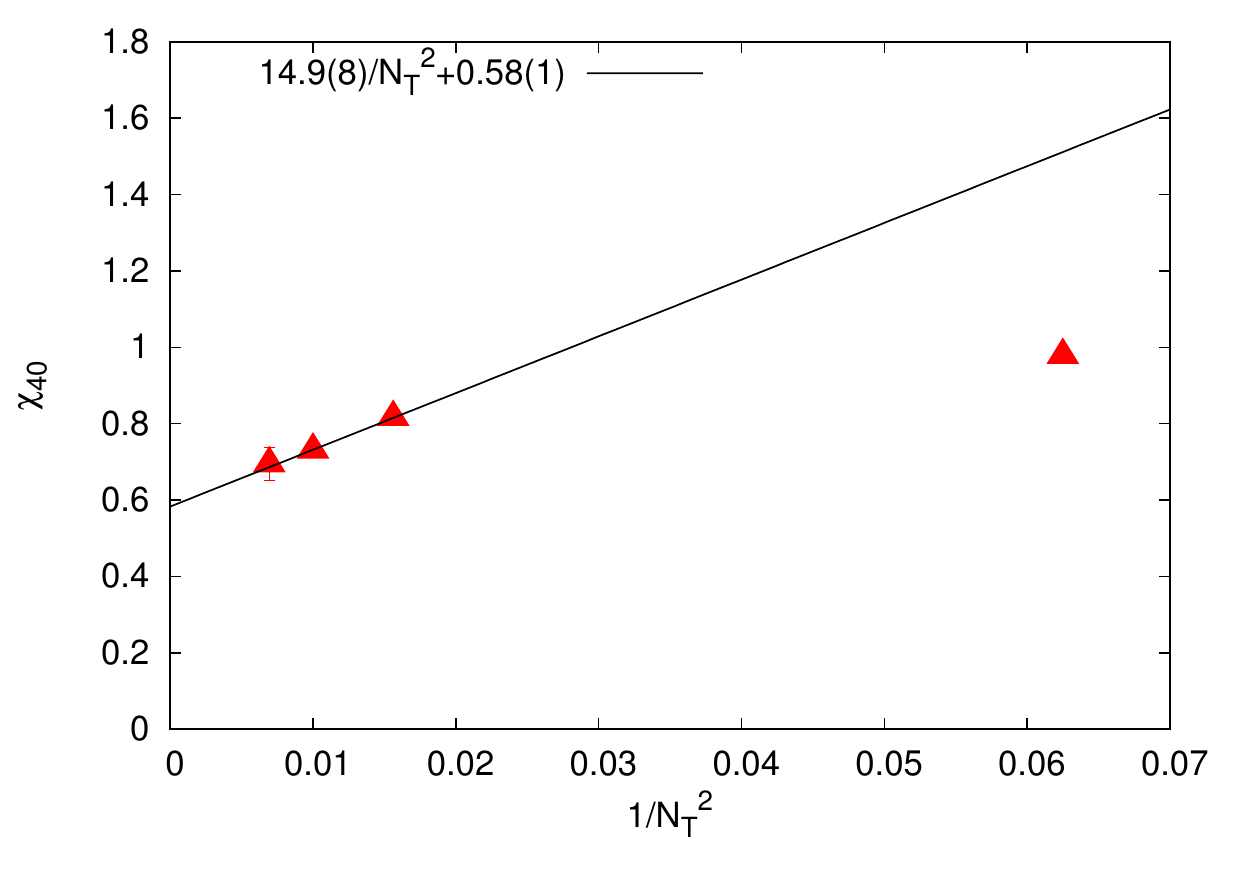}
\includegraphics[scale=0.425]{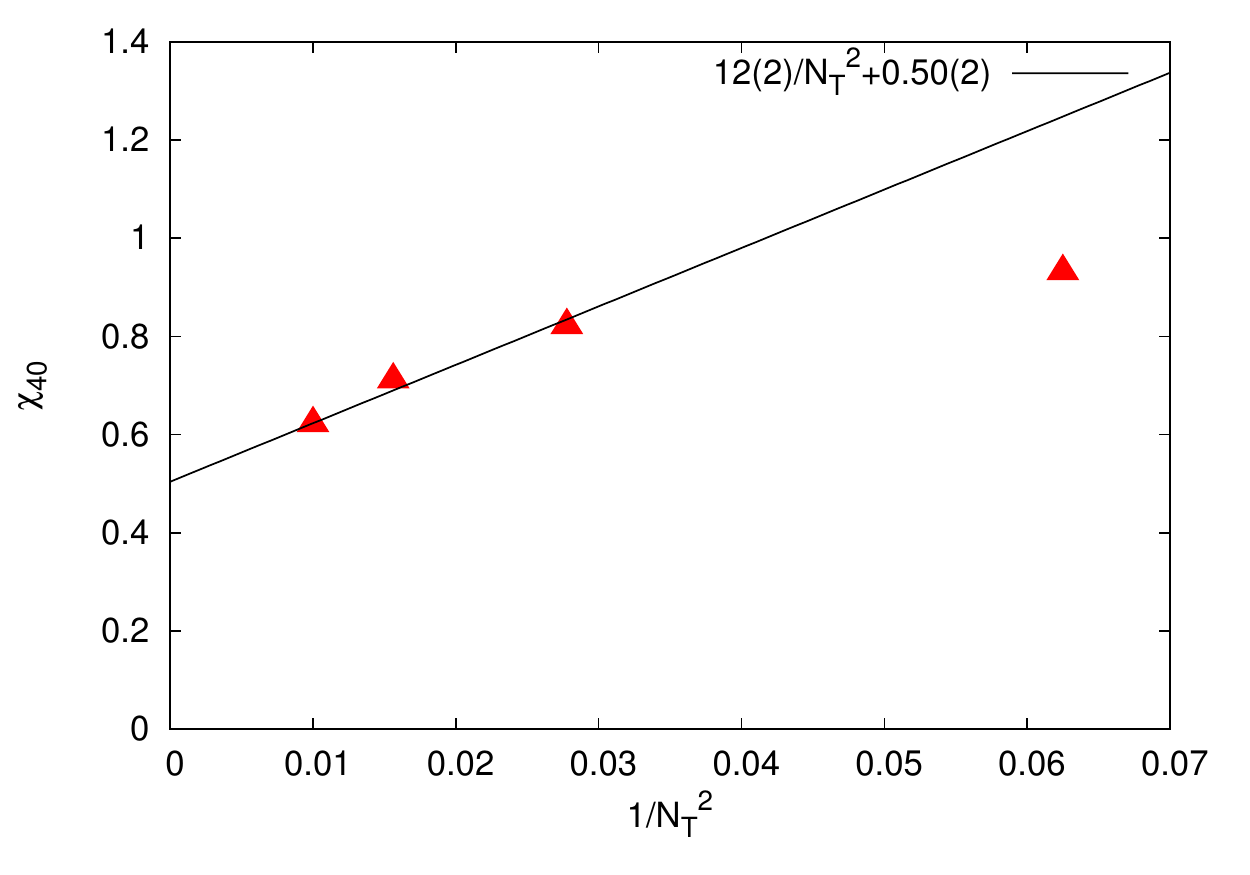}
\caption{The continuum extrapolation for the fourth order diagonal
susceptibility in quenched QCD at $1.25 T_c$ (left panel) and $2 T_c$ (right
panel) as a function of the lattice spacing or equivalently $N_t$ for
$m/T_c=0.1$. } \label{fig:chi4quench1}
\end{center}
\end{figure} 

\section{Summary} 

In two recent studies~\cite{gs14,gs12}, we revisited the idea of introducing
$\mu$ as a linear ($\mu N$) term on the lattice, where $N$ is the conserved
charge on the lattice.  It is known to lead to $\mu$-dependent quadratic
divergences in the number/energy density. Understanding the nature and origin of
these divergences is both theoretically important as well as computationally
advantageous since a suitable method of their elimination can allow for its
much wider applicability including for exact chiral fermions.

We show here that the divergent $1/a^2$ term in the second order susceptibility
for free fermions is not a mere lattice artifact but it is present even in the
continuum.  Following the practice there, it can be subtracted off on the
lattice.  Since $N$ is a conserved charge, it does not get renormalized in the
interacting theory. This implies that the coefficients of the divergent $1/a^2$
term in full QCD should be the same as for the free fermions.  We show that
subtraction of such a free fermion term from the susceptibility in the
(quenched) interacting case leaves only a physical piece.  It has no divergences
in the continuum limit and its extrapolation to continuum limit agrees with that
computed by analytically eliminating the divergence. As in perturbation
theory,  once the zero temperature divergence is removed, interactions do not
induce any additional divergence at finite $T$ or $\mu$ holds true non-perturbatively as well.  We conclude that actions linear in $\mu$ can
be employed safely with added computational advantages in computing the known
fourth and higher order quark number susceptibilities, as well as yet higher
order ones.

\end{document}